\documentclass[12pt]{iopart}
\usepackage{psfig}
\usepackage{graphicx}
\usepackage{graphics}
\begin{document}

\title{Excitation Function of the Longitudinal Expansion in Central Nuclear Collisions}

\author{Marcus Bleicher}

\address{Institut f\"ur Theoretische Physik,
J.~W.~Goethe Universit\"at, 60054 Frankfurt am Main,
Germany}

\begin{abstract}
Longitudinal hadron spectra from Proton-Proton (pp) and
nucleus-nucleus (AA) collisions from $E_{lab}= 2$~AGeV to $\sqrt
s=200$~AGeV are investigated. The widths of the rapidity spectra for
various particle species increases monotonously with energy. The
present calculation indicates no sign of a step like behaviour as
excepted from the Kaon transverse mass systematics. For Pions,
the transport simulation is consistent with a Landau type scaling of the
rapidity widths, both in central AA reactions and in pp collisions.
However, other hadron species do not follow the Landau scaling. The
present model predicts a decreasing rapidity width with particle
mass for newly produced particles, not supporting a Landau type flow interpretation.
\end{abstract}

\maketitle\nopagebreak

Based on recent lattice QCD (lQCD) calculations it has been
speculated that partonic degrees of freedom might already lead to
visible effects at $\sim$ 5 $A\cdot$GeV \cite{MT-prl}. Especially the
hardening of the measured transverse mass ($m_t$) spectra in central
Au+Au collisions relative to pp interactions \cite{NA49_T,Goren}
around AGS energies obtained great interest and was studied in
detail \cite{Bratkovskaya:2004kv}. This increase of the inverse
slope parameter $T$ is commonly attributed to strong collective
flow, which is absent in the respective pp or pA collisions. It has also
been proposed \cite{SMES} to interpret the high and approximately
constant $K^\pm$ slopes above $\sim 30$ AGeV -- the 'step' -- as an
indication of the phase transition.

\noindent
This interpretation seems supported by microscopic
transport simulations
\begin{itemize}
\item
indicating the increasing importance of sub-hadronic degrees of
freedom above AGS energies \cite{Weber98} and
\item
from the comparison of the thermodynamic parameters $T$ and $\mu_B$
extracted from the transport models in the central overlap region
\cite{Bravina} with the experimental systematics on chemical
freeze-out configurations \cite{Braun-Munzinger:1996mq,Braun-Munzinger:1998cg,Cleymans} in the $T,\mu_B$ plane.
\end{itemize}

Let us now explore whether a similar 'step' is also present in the
excitation function of longitudinal observables.
For the present study we employ the UrQMD model (v2.2)
\cite{UrQMD1,UrQMD2}. It takes into
account the formation and multiple rescattering of hadrons and
dynamically describes the generation of pressure in the hadronic
expansion phase. It involves also interactions of (di-)quarks, however  
gluonic degrees of freedom are not treated explicitly, but are implicitly present  in strings.
This simplified treatment is generally accepted to describe Proton-Proton and Proton-nucleus
interactions. 
\begin{figure}
\begin{tabular*}{17cm}{ll}
  \psfig{file=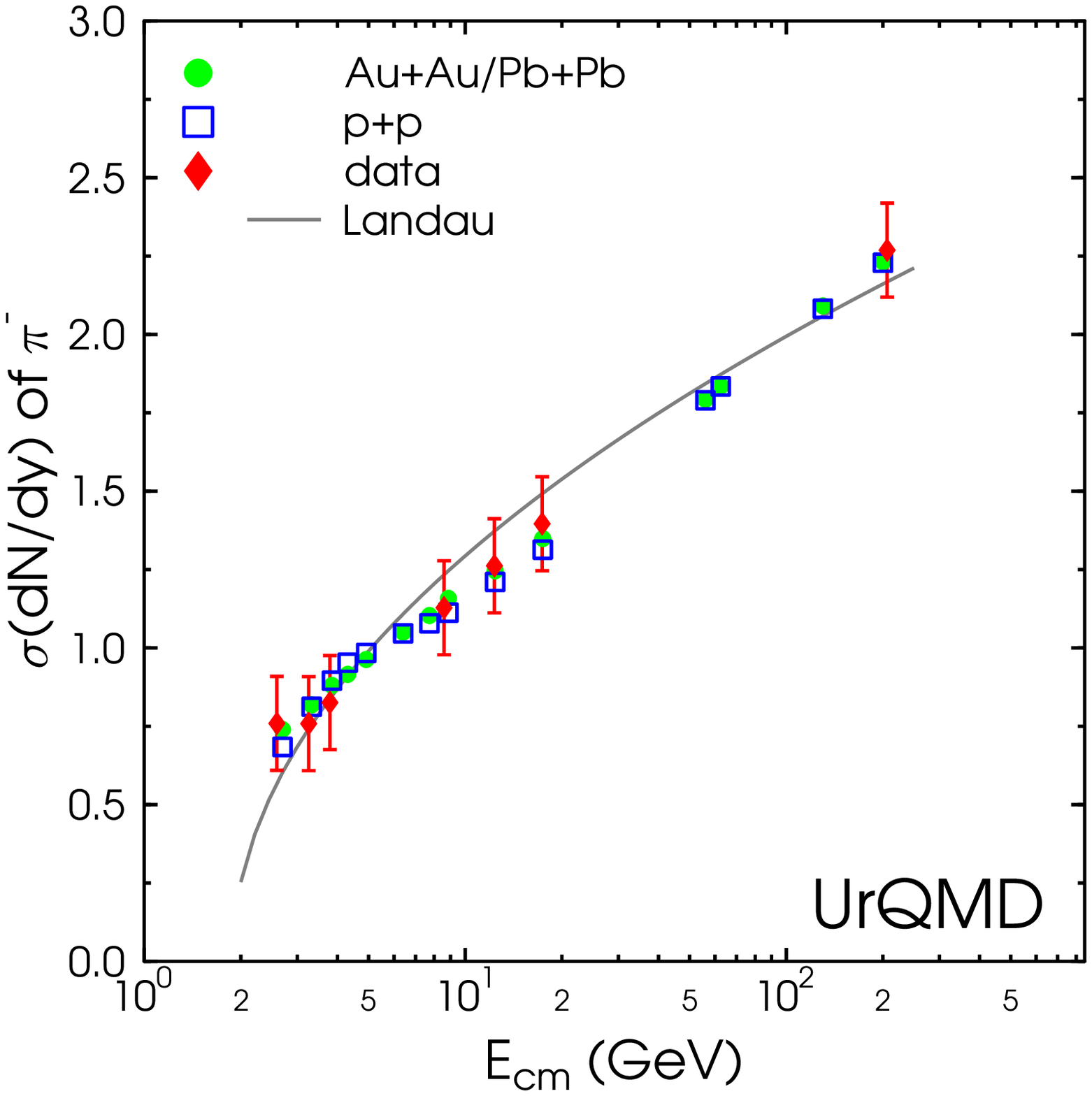,width=8cm} & \psfig{file=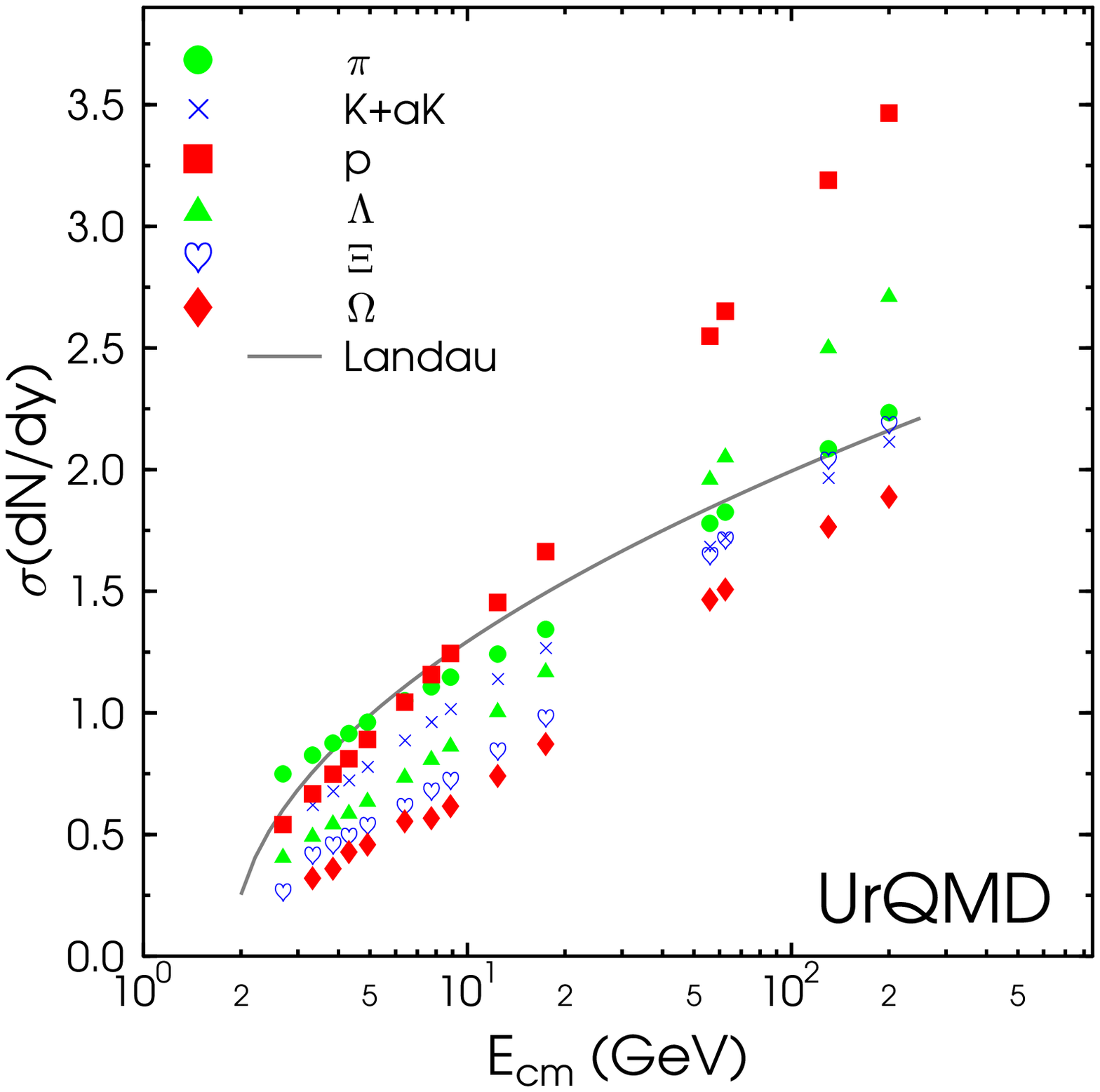,width=8cm} \\
\end{tabular*}\vspace*{-0.3cm}
\caption{Left: \label{rapwidth} The root mean square of the rapidity
distribution of negatively charged Pions in central Au+Au/Pb+Pb and
Proton+Proton reactions as a function of the center of mass energy.
UrQMD calculations for Au+Au/Pb+Pb are denoted by full circles, the
pp results are shown by open squares. The prediction from Landau's
model is given by the line (Eq. \ref{eq1}). Data \cite{Roland:2004}
are depicted by full diamonds. Right: \label{particles1} The root
mean square of the rapidity distribution of Pions, Kaons, Protons,
Lambdas, Cascades and Omega hadrons in central Au+Au/Pb+Pb reactions
as a function of center of mass energy. UrQMD calculations for
Au+Au/Pb+Pb are denoted by symbols. The prediction from  Landau's
model is given by the line.}
\vspace*{-0.6cm}
\end{figure}

 It became popular
to interpret relativistic heavy ion reactions with Landau's
hydrodynamical model
\cite{Fermi:1950jd,Landau:gs,Belenkij:cd,Carruthers:ws,Carruthers:dw,Carruthers:1981vs}
(for recent applications of this model to relativistic
nucleus-nucleus interactions see
\cite{Stachel:1989pa,Steinberg:2004vy,Murray:2004gh,Roland:2004}). Therefore we
will use this simple hydrodynamical picture as a baseline for the
model and data comparison. The main physics assumptions of Landau's
picture are: The collision of two Lorentz-contracted hadrons or
nuclei leads to full thermalization in a volume of size
$V{m_p}/\sqrt{s}$. This justifies the use of thermodynamics and
establishes the system size and energy dependence. A simple equation
of state  $p=\epsilon/3$ is assumed. Chemical potentials are usually
assumed to vanish. The main results derived from these assumptions
are: A universal formula for the produced entropy, determined mainly
by the initial Lorentz contraction and Gaussian rapidity distributions,
at least for newly produced particles. The results can be summarised
in the energy dependent rapidity density \cite{Carruthers:dw}:
\begin{equation}
\frac{dN}{dy}=\frac{Ks^{1/4}}{\sqrt{2\pi L}}\exp(-\frac{y^2}{2L})\quad {\rm with}\quad L=\sigma^2_y={\rm ln}(\sqrt s/2m_p).
\label{eq1}
\end{equation}

As depicted in Fig. \ref{rapwidth} (left) the UrQMD predictions (full circles) for the rapidity widths 
of negatively charged Pions 
in Au+Au (Pb+Pb) reactions are in line with the experimental data \cite{Roland:2004} (full diamonds) 
and  Landau's hydrodynamical model (full line).
A rather surprising observation is that the calculated rapidity widths of $\pi^-$ in pp interactions 
(open squares) are identical to the AA results.

The rhs. of Fig. \ref{rapwidth} shows the rapidity widths of different particle species as a function of
energy. Here the calculation (for hadrons other than Pions) differs considerably from the
Landau model: (I) Hadrons containing (initial) up or down quarks show a strong increase of the
rapidity widths with energy (leading particle effect). (II) Hadrons without initial up or downs quarks
show a decreasing rapidity width with increasing mass at fixed energy.  

\begin{figure}[t]
\begin{minipage} [r] {6cm}
\psfig{file=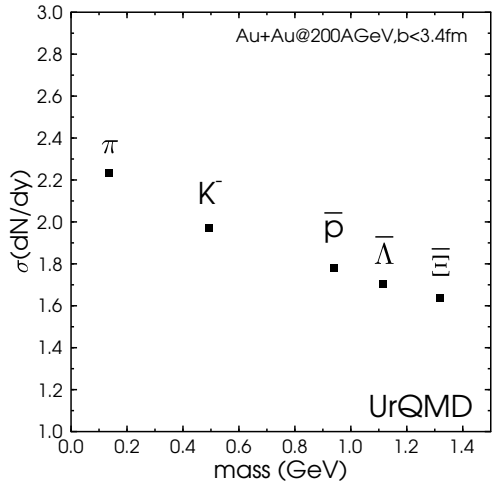,width=8cm}
\end{minipage}
\hspace*{-2cm}\begin{minipage} [r] {9cm}
\caption{\label{mass} Root mean
square of the rapidity distribution as a function of particle mass
in central Au+Au  reactions at $\sqrt s=200$~AGeV.}
\end{minipage}
\vspace*{-1.2cm}
\end{figure}

The second feature is  shown in detail in Fig. \ref{mass} for central Au+Au reactions at $\sqrt s=200$~AGeV. 
Here the root mean square of hadrons without initial quarks is given as a function of particle
mass. Beginning from Pions with rapidity width of 2.2 units up to anti-Xi baryons with a rapidity width of
only 1.65 units. Within 10\% deviations, this results in Bjorken-plateau of $y_{cm}\pm 0.75$.

In conclusion, we have explored the excitation functions of the rapidity widths of
Pions, Kaons, Protons, Lambdas and Xis in pp and/or AA collisions.
The rapidity spectra of Pions for all investigated systems and energies can be described well by Gaussians.
The energy dependence of the width of the Pion rapidity distribution follows the
prediction of  Landau's hydrodynamical model, both in nucleus-nucleus and Proton-Proton interactions.
In nucleus-nucleus reactions, the width of all other investigated hadrons deviates from the Landau picture.
For newly produced hadrons, the present calculation shows a strong mass dependence of the
rapidity width, allowing for a Bjorken-plateau only in a narrow window ($y_{cm}\pm 0.75$ at the highest 
RHIC energy) around midrapidity.
We do not observe any irregularities (e.g. steps or peaks) in the rapidity width of all investigated
hadrons. 

\noindent
\section*{Acknowledgements}
This work was supported by GSI, DFG and BMBF.
This work used computational resources provided by the
Center for Scientific Computing at Frankfurt (CSC).


\section*{References}
\begin{thebibliography}{10}

\bibitem{MT-prl}
    E. L. Bratkovskaya {\it et al.},
    Phys. Rev. Lett. {\bf 92}, 032302 (2004)

\bibitem{NA49_T}
    V. Friese {\it et al.}, NA49 Collaboration,
    J. Phys. G {\bf 30}, 119 (2004).

\bibitem{Goren}
    M. I. Gorenstein, M. Ga\'zdzicki, and K. Bugaev,
    Phys. Lett. B {\bf 567}, 175 (2003).

\bibitem{Bratkovskaya:2004kv}
E.~L.~Bratkovskaya {\it et al.},
Phys.\ Rev.\ C {\bf 69}, 054907 (2004)

\bibitem{SMES}
    M. Gazdzicki and M. I. Gorenstein,
    Acta Phys. Polon. B {\bf 30}, 2705 (1999).

\bibitem{Weber98}
    H. Weber, C. Ernst, M. Bleicher {\it et al.},
    Phys. Lett. B {\bf 442}, 443 (1998).

\bibitem{Bravina}
    L. V. Bravina {\it et al.}, Phys. Rev. C {\bf 60}, 024904 (1999),
     Nucl. Phys. A {\bf 698}, 383 (2002).

\bibitem{Braun-Munzinger:1996mq}
  P.~Braun-Munzinger and J.~Stachel,
  Nucl.\ Phys.\ A {\bf 606}, 320 (1996).

\bibitem{Braun-Munzinger:1998cg}
  P.~Braun-Munzinger and J.~Stachel,
  Nucl.\ Phys.\ A {\bf 638}, 3 (1998).

\bibitem{Cleymans}
    J. Cleymans and K. Redlich, Phys. Rev. C {\bf 60}, 054908 (1999).

\bibitem{UrQMD1}
    S.A.~Bass {\it et al.}, Prog. Part. Nucl. Phys. {\bf 42}, 255 (1998).

\bibitem{UrQMD2}
    M.~Bleicher {\it et al.},  J. Phys. G {\bf 25}, 1859 (1999).

\bibitem{Fermi:1950jd} E.~Fermi, Prog.\ Theor.\ Phys.\  {\bf 5}, 570 (1950).

\bibitem{Landau:gs} L.~D.~Landau, Izv.\ Akad.\ Nauk Ser.\ Fiz.\  {\bf 17}, 51 (1953).

\bibitem{Belenkij:cd} S.~Z.~Belenkij and L.~D.~Landau, Usp.\ Fiz.\ Nauk {\bf 56}, 309 (1955).

\bibitem{Carruthers:ws} P.~Carruthers and M.~Doung-van, Phys.\ Rev.\ D {\bf 8}, 859 (1973).

\bibitem{Carruthers:dw} P.~Carruthers, Annals N.Y.Acad.Sci. 229, 91 (1974).

\bibitem{Carruthers:1981vs} P.~A.~Carruthers, LA-UR-81-2221 

\bibitem{Stachel:1989pa}
  J.~Stachel and P.~Braun-Munzinger,
  Phys.\ Lett.\ B {\bf 216},  1 (1989).

\bibitem{Steinberg:2004vy}
P.~Steinberg,
arXiv:nucl-ex/0405022.

\bibitem{Murray:2004gh} M.~Murray, arXiv:nucl-ex/0404007.

\bibitem{Roland:2004} G. Roland, Quark Matter 2004.

\end{thebibliography}
\end{document}